\documentclass[reqno,12pt,a4paper]{amsart}
\usepackage[english]{babel}
\usepackage[T1]{fontenc}
\usepackage{mathtools,amssymb,verbatim,dsfont}
\usepackage{physics}
\usepackage{tikz}
\usepackage{tikz-3dplot}
\usepackage{pgfplots}
\usetikzlibrary{babel,3d}
\usepackage{bm}
\usepackage{float}
\usepackage{enumerate}
\usepackage{xfrac}
\usepackage{csquotes}
\usepackage[hidelinks]{hyperref}
\usepackage{caption}
\usepackage[backend=biber, style=numeric, sorting=nty]{biblatex}
\usepackage{etoolbox}
\usepackage{graphicx}
\usepackage{algorithm}
\usepackage{algpseudocode}
\usepackage{lineno}
\usepackage{fancyhdr}
\usepackage{subcaption}
\usepackage{cleveref}
\usepackage{pdfpages}
\usepackage{xcolor}
\usepackage{bigfoot} 
\usepackage{chngcntr}
\usepackage{authoraftertitle}
\usepackage{multirow}

\newtheorem{theorem}{Theorem}[section]

\newtheorem{conjecture}[theorem]{Conjecture}
\theoremstyle{definition}

\theoremstyle{remark}
\newtheorem{remark}[theorem]{Remark}

\numberwithin{equation}{section}

\DeclareMathOperator{\id}{id}

\DeclareMathOperator{\pr}{pr}

\DeclareMathOperator{\Cov}{Cov}

\DeclareMathOperator{\dist}{dist}
\DeclareMathOperator{\vol}{dVol}

\DeclareMathOperator{\eqdist}{\overset{d}{=}}

\DeclareMathOperator{\St}{St}

\title{Extended Kalman Filtering on Stiefel Manifolds}

\author{Jordi-Llu\'{i}s Figueras
\and
Aron Persson
\and
Lauri Viitasaari}

\makeatletter
\let\newtitle\@title
\let\newauthor\@author
\makeatother

\pgfplotsset{compat=1.18} 
\begin{document}

\begin{abstract}
    A generalisation of the extended Kalman filter for Stiefel manifold-valued measurements is presented.  We provide simulations on the 2-sphere and the space of orthogonal 4-by-2 matrices which show significant improvement of the Extended Kalman Filter compared to only relying on raw measurements.
\end{abstract}

\maketitle
\thispagestyle{empty}

\tableofcontents

\section{Introduction}

\subsection{Examples of projected dynamics}

There are multiple situations where projected dynamics is inherently tied to gaining information. We shall begin by giving simple examples of two such situations in order to broaden the perspective before we go into the specifics relating to applications of Stiefel manifold-valued statistics in radiology.

Consider a radioactive dust particle travelling through an open space. Suppose also there is a moderate wind going through the area and we can only measure the direction of the incoming radiation. In physics one often models the movement of dust particles as if they travel like Brownian motion with a drift (the drift here comes from the wind). Therefore the dust particle does not travel deterministically and there is an uncertainty inherent to the position of the dust particle. Moreover, the measured radiation incoming into our sensor might have a somewhat perturbed inclination error as it collides with air particles in between the dust particle and the sensor. Hence, there is an uncertainty of the incoming direction of the radiated particles as well. In this situation the dust particle is travelling across $\mathbb{R}^3$ and our measurements are taken on the sphere $\mathbb{S}^2$. The information obtained is (non-linearly) projected onto the sphere.

Another plausible scenario would be a circular particle accelerator. As a charged particle is centripetally accelerated through electro-magnetic forces we measure its position in the accelerator using sensors throughout the walls of the accelerator. As the particle travelling through the accelerator is a quantum particle, the position of the particle is inherently probabilistic. At the same time the sensors are not perfect and there is uncertainty in the measurements. Here the particle moves inside a tubular neighbourhood of a circle and the measurements happen on the torus $\mathbb{T}^2$.

In MIMO radio-systems for cellular networks a configuration of $n$ antennas interfere constructively and a directed radio signal is produced with high energy efficiency. This signal is then received at and responded by $k$ receivers. These receivers move around, in part, in a predictive way (a person moving in a certain direction will probably move in that direction for a while) and, in part, with some uncertainty. Hence, these receivers may be modelled as stochastic processes with some dynamical component.  Moreover, the antenna configurations can be modelled as being a vector in $\mathbb{R}^n$ for each receiver. In total we consider the state as an $n$ by $k$ matrix. However, as these configurations do not depend on the total strength of the currents, only their relative strength and phase, so measurements are realised as $n$ by $k$ orthogonal matrices, see \cites{hussien2014multi,pitaval2013joint,seddik2017multi}. These realisations may be computed by only retaining the orthogonal part, $Q$, in the polar decomposition of the $n$ by $k$ matrix. This $Q$ can be computed so that it is a projection onto the the space of $n$ by $k$ orthogonal matrices, i.e. the Stiefel manifold $\St_{n,k}$. These measurements are assumed to be noisy and the measured optimal configuration of the antennae  has some uncertainty.

All these three situations illuminates a mathematical problem: an object moves around with a drift in some (possibly an open subset of) vector space, see Figure \ref{fig:measurementsonmanifold} for a picture of these situations. If one performs measurements on a non-linear space, i.e. if $x,y\in X$, then it is generally not true that $x+y\in X$. How would one filter out this information using both the known drift of the system and the measurements? In all the above examples the non-linear spaces are manifolds which have sufficient geometrical structure which allows for measurement of distances between points. Using distances and curves on these manifolds it is then possible to weight the prediction together with the measurement. In a nutshell, working on the original filtering equations developed in \cites{kalman1960new,kalman1960contributions,kalman1961new} we will filter the predicted point and the measurement by using a curve that connects these two points. Then one can obtain the filtered mean by taking a weighted average along the curve.

This paper can be viewed as part of the vast field of directional statistics. The area of directional statistics has found applications in medicine, see \cites{craig2010different,demir2019application,karaibrahimoglu2021circular}, in meteorology, see \cites{li2018wavefront,nunez2015bayesian,Nunez-Gutierrez-Escarela-2011}, and in robotics, see \cite{lang2012mpg,sveier2018pose}, to name a few.

\subsection{Organisation of the paper}

Section \ref{sec:background} explains the foundational concepts for the problem at hand. In Section \ref{sec:kalman} we give an exposition on how to use the extended Kalman filter when the measurements are points on the Stiefel manifold $\St_{n,k}$. The reader with little interest of the geometrical meaning behind the Extended Kalman filter on $\St_{n,k}$ may jump directly to Algorithm \ref{alg:kalmansphere} at the end of this section. In Sections \ref{sec:SimS2} and \ref{sec:simSt} we give results from simulations for the 2-sphere and $\St_{4,2}$, respectively. In Section \ref{sec:discussion} we discuss some limitations and we present some potential drawbacks of the extended Kalman filter on Stiefel manifolds.

\section{Setting}
\label{sec:background}

Suppose an object $X_t$ is represented as an $n$ by $k$ matrix, that is $X_t \in \mathbb{R}^{n\times k}$. As $t$ increases we suppose it moves around with known dynamics and with non-zero noise (therefore the movement is not fully deterministic). This may be mathematically represented as an SDE (Stochastic Differential Equation)
\begin{equation}
\dd X_t = A X_t \dd t + \nu \dd B_t, \qquad X_0 \eqdist N(\mu_0,\nu_0^2 \id_{\mathbb{R}^{n\times k}}).
\label{eq:systemsde}
\end{equation}
    Here $\id_{\mathbb{R}^{n\times k}}$ denotes the identity map from the space of $n$ by $k$ matrices to itself, i.e. for given $Y\in \mathbb{R}^{n\times k}$, $\id_{\mathbb{R}^{n\times k}}(Y)=Y$. More explicitly, if $Y_{\text{vec}}\in \mathbb{R}^{nk}$ is the vectorisation of an $n$ times $k$ matrix $Y$, then if $\id_{\mathbb{R}^{n\times k}}^{\text{vec}}$ is the corresponding vectorized linear map, it is the $nk \times nk$ identity matrix.
The initial value $X_0$ is the starting (normal) distribution of $X_t$ and $A\in \mathbb{R}^{n\times n}$ describes the dynamics which is the infinitesimal displacement over time. The term $\dd B_t$ is a formal stochastic differential of the Brownian motion which we scale by a factor $\nu$. After a time step $\delta t>0$ and if $A=0$, then solving \eqref{eq:systemsde} one would obtain $X_t=X_0+\nu \int_0^{\delta t} \dd B_t$ which has variance matrix $(\nu_0^2+\nu^2 \delta t) \id_{\mathbb{R}^{n\times k}}$ and constant mean $\mu_0$. More generally, if $A$ is anti-symmetric, i.e. $A^T = -A$, then one obtains the full solution to \eqref{eq:systemsde} as
\[
X_t\eqdist N(\exp_{\mathrm{M}}(tA) \mu_0,(\nu_0^2 + t \nu^2)\id_{\mathbb{R}^{n\times k}}).
\]
Here $\exp_{\mathrm{M}}:\mathbb{R}^{n\times n}\rightarrow \mathbb{R}^{n\times n}$ denotes the matrix exponential defined as
\[
\exp_{\mathrm{M}}(A) = \sum_{j=0}^\infty \frac{A^j}{j!},
\]
where $A^j$ denotes $j$-times matrix multiplication and $A^0=I_{n\times n}$, the $n \times n$ identity matrix.
(Recall that this series is absolutely convergent for any $A$).

\begin{remark}
Note that this solution is only true when the factor $\nu$ is a scalar (for which we then call the noise isotropic). If the factor $\nu$ is not a scalar then there will appear more time dependent terms in the covariance part.    
\end{remark}

\begin{figure}[ht!]
    \centering
    \includegraphics[width=0.9\textwidth]{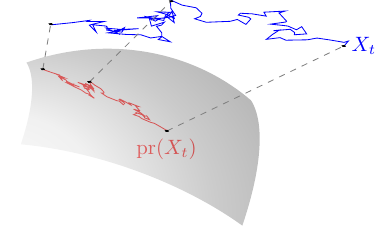}
    \caption{A picture of a stochastic process which is then non-linearly projected onto a non-linear space.}
    \label{fig:projectionontomanifold}
\end{figure}

Suppose now the process $X_t$ from \eqref{eq:systemsde} is observed, but we only observe the orthogonal part in the polar decomposition of $X_t$. This observation is a projection onto the space of $n\times k$ matrices with pairwise orthonormal columns, the Stiefel manifold 
\[
\St_{n,k}:= \left\{X \in \mathbb{R}^{n\times k}: X^T X = I_k \right\}.
\]
We write $\pr:\mathbb{R}^{n\times k}\rightarrow \St_{n,k}$, and its computation can be done by a polar decomposition using SVD algorithms. If $X_t =Q_t D_t S_t^*$ is the thin SVD-decomposition, see \cite{bai2000templates}, where $D_t$ and $S_t$ are $k\times k$  matrices and $Q_t$ is $n \times k$, then $\pr(X_t)=Q_t S_t^*$. Note that this map is heavily non-linear and should not be confused with (linear) projections seen in linear algebra. A conceptual picture on how this projection of the stochastic process $X_t$ might look like is shown in Figure \ref{fig:projectionontomanifold}.

\begin{figure}[ht!]
    \centering
    \includegraphics[width=0.9\textwidth]{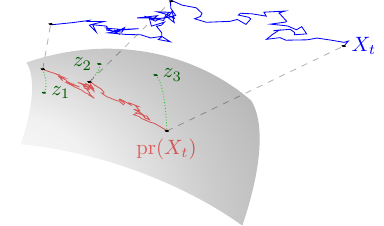}
    \caption{A picture of a stochastic process $X_t$ that is (non-linearly) projected onto a manifold together with some measurements $z_1,z_2,z_3$} 
    \label{fig:measurementsonmanifold}
\end{figure}

Suppose further that when this projected process is observed there is an error of magnitude $\xi^2$ inherent to these measurements. That is, if $\pr(X_t)$ is the orthogonal part of $X_t$ using the polar decomposition, the measurement's error of $X_t$ is assumed to have mean $\pr(X_t)$ and (scalar) variance $\xi^2$ (we shall delve into what this means in the next section). In Figure \ref{fig:measurementsonmanifold} there is an illustration of how these faulty measurements may look like.

\begin{figure}[ht!]
    \centering
    \includegraphics[width=0.9\textwidth]{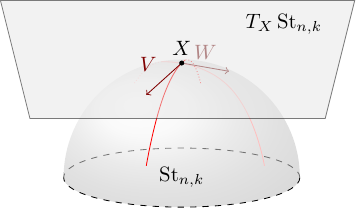}
    \caption{A conceptual picture of the tangent space at a point $X\in \St_{n,k}$, together with two tangent vectors $V,W\in T_{X} \St_{n,k}$ and their corresponding curves that has the corresponding momentous velocity at $X$.}
    \label{fig:tangent}
\end{figure}

Now the goal is to estimate $\pr(X_t)$ given that we know the system parameters $A\in \mathbb{R}^{n\times n}$, $\nu \in \mathbb{R}$ and the measurement error. In order to capture parts of this non-linearity we shall use the tangent space. The tangent space $T_X \St_{n,k}$ contains all possible velocities (aka. tangent vectors) at the point $X\in \St_{n,k}$ such that any curve going through $X$ with velocity in $T_X \St_{n,k}$ will (infinitesimally) stay on $\St_{n,k}$, see Figure \ref{fig:tangent} for an illustration. Explicitly, the tangent space is given by
\[
T_{X} \St_{n,k}= \left\{ V\in \mathbb{R}^{n\times k}: V^TX=-X^TV \right\}
\]
which may be deduced from that any curve $Y(t)$ on $\St_{n,k}$ must satisfy $Y^T(t) Y(t)= I_{k}$. Differentiating this expression yields exactly the above set of tangent vectors.

\begin{figure}[ht!]
    \centering
    \includegraphics[width=0.9\textwidth]{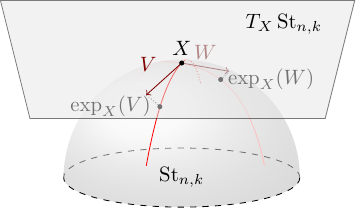}
    \caption{An illustration of the exponential map and how it maps tangent vectors $V,W\in T_X\St_{n,k}$ to points on $\St_{n,k}$.}
    \label{fig:exponential}
\end{figure}

Now, considering $\St_{n,k}$ as a subset\footnote{This is not an isometric embedding into $\mathbb{R}^{n\times k}$ when equipped with the Frobenius norm, see Remark~\ref{rem: frobenius norm}.} of $\mathbb{R}^{n\times k}$, one can measure distances on $\St_{n,k}$ by considering the length of curves on $\St_{n,k}$ with constant velocity by restricting a modified version of the Frobenius norm onto the tangent space at every point. Recall that the classical Frobenius norm of a matrix $B\in \mathbb{R}^{n\times k}$ is defined by $\norm{B}^2=\trace(B^TB)$. Through the notion of distance, one can relate points on the tangent space $T_X\St_{n,k}$ to points near $X$: by denoting $\exp_X(V)$ as the point $Y\in \St_{n,k}$ which is reached by taking the constant velocity curve starting at $X$ with starting velocity direction $V/\norm{V}$ and is the distance $\norm{V}$ away. Note here that $\exp_X$ is not the same as the matrix exponential map $\exp_{\mathrm{M}}$ defined earlier. The map $\exp_X:T_X \St_{n,k} \rightarrow \St_{n,k}$ is called the (Riemannian) exponential map (we have provided a picture of this exponential map in Figure \ref{fig:exponential}). We denote the inverse of $\exp_X$ by $\log_X$, and call it the logarithm map.

\begin{remark}
    Explicitly, one may compute $\exp_X(V)$ by first computing the QR-decomp\-osition
    \[
    QR= (I_{n\times n}-X X^T)V
    \]
    (here $Q$ is a $n$ by $k$ orthogonal matrix and $R$ is a $k$ by $k$ upper triangular matrix), then
    \[
    \exp_X(V) = \begin{pmatrix}
        X & Q
    \end{pmatrix} \exp_{\mathrm{M}}\left(\begin{pmatrix}
        X^T V &- R^T\\ R & 0_{k\times k}
    \end{pmatrix} \right) \begin{pmatrix}
        I_{k\times k}\\ 0_{k\times k}
    \end{pmatrix}.
    \]
    In Theorem 2.1 and Corollary 2.2 in \cite{edelman1998geometry} one can find more information about this exponential map. To compute the logarithm, which is a little bit more involved to compute but still possible, we refer the reader to \cite{zimmermann2017matrix}.
\end{remark}

The inner product, aka metric, on the tangent space $T_X \St_{n,k}$, with $X\in \St_{n,k}$, is computed as follows
\begin{equation}
    \langle V,W\rangle_X = \trace\left( V^T (I_{n\times n} -\frac{1}{2} XX^T)W) \right).
    \label{eq:stmetric}
\end{equation}

\begin{remark}\label{rem: frobenius norm}
An important key point to note is that $\St_{n,k}$ is not considered as a subset of $\mathbb{R}^{n \times k}$ with the standard Frobenius norm. Here, the norm induced by \eqref{eq:stmetric} is the norm inherited from the Frobenius norm on the Lie group $O(n)$ of which the Stiefel manifold is a quotient space. This actually makes $\St_{n,k}$ a isometrically submerged submanifold of $O(n)$ and by extension $\mathbb{R}^{n\times n}$.  It is still an open question how and if one can isometrically embed $\St_{n,k}$ in $\mathbb{R}^{n \times k}$ with an easily expressed norm defined on $\mathbb{R}^{n\times k}$.
\end{remark}

\begin{figure}[ht!]
    \centering
    \includegraphics[width=0.9\textwidth]{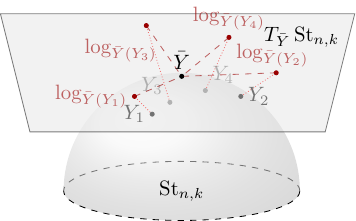}
    \caption{A picture on how the inherent average $\bar{Y}$ relates to a set of points $Y_1,Y_2,Y_3,Y_4\in \St_{n,k}$.}
    \label{fig:average}
\end{figure}

Through the logarithm map one can define a notion of average which is inherent (aka. intrinsic) to $\St_{n,k}$. Given points $Y_1,\dots,Y_N\in\St_{n,k}$, their average is defined as $\bar{Y}\in\St_{n,k}$ which minimises the distance squared from all the points $Y_1,\dots,Y_N$. Locally, the average $\bar{Y}\in\St_{n,k}$ explicitly satisfies the barycenter equation 
\[
\sum_{i=1}^N \log_{\bar{Y}}(Y_i)=0.
\]
This average is illustrated in Figure \ref{fig:average}. Note that this is a non-linear generalisation of the standard definition of average in a vector space: in this case the average $\bar{y}\in \mathbb{R}$ satisfies
\[
\frac{1}{N}\sum_{i=1}^N (y_i - \bar{y})=0
\]
which is a bit more convoluted way of saying
\[
\bar{y}= \frac{1}{N}\sum_{i=1}^N y_i.
\]

Let $B_j$ be an orthonormal basis for the vector space $T_{\bar{Y}} \St_{n,k}$ under the inner product \eqref{eq:stmetric}. The sample covariance given points $Y_1,\dots,Y_N\in\St_{n,k}$, is then the $nk-\frac{1}{2}k(k+1)$ by $nk-\frac{1}{2}k(k+1)$ matrix
\[
(\Cov_Y)_{i,j}= \frac{1}{N-1} \sum_{\ell=1}^N \langle \log_{\bar{Y}}(Y_\ell),B_i\rangle_{\bar{Y}} \langle \log_{\bar{Y}}(Y_\ell),B_j\rangle_{\bar{Y}}
\]
with corresponding scalar variance
\begin{equation}
    \frac{\trace(\Cov_Y)}{nk-\frac{1}{2}k(k+1)}.
    \label{eq:scalarcariancesample}
\end{equation}
Note also that the notion of variance has a meaning of measuring the squared distance, and the mean has the meaning of being the point which minimises the least squared distance, see \cite[Definition 2.1 and 2.2]{figueras2024parameter}.

Observe that spheres  $\mathbb{S}^{n-1}$ are special cases of $\St_{n,k}$. Specifically note that $\mathbb{S}^{n-1}$ is the same as $\St_{n,1}$. On the spheres, we know that given a normal random variable $X$ in $\mathbb{R}^{n+1}$ with mean $\mu$ and isotropic covariance $\sigma^2 I_{n+1}$ the projected random variable $\pr(X)=\frac{X}{\norm{X}}$ has inherent average $\frac{\mu}{\|\mu\|}$. Moreover, it has been shown that there is unique, one-to-one, intrinsic scalar variance of $\pr(X)$ for every choice of $\sigma$ for $\mathbb{S}^2$, see \cite[Theorem 2.8]{figueras2024parameter}. Along this paper we shall assume that the same is true for all $\St_{n,k}$.

Let $X\in \mathbb{R}^{n\times k}$ is a normal random variable with mean $\mathbb{E}[X]\in \St_{n,k}$ and covariance matrix $\Cov(X)=\sigma^2 \id_{\mathbb{R}^{n\times k}}$. Then the inherent scalar variance can be computed by
\begin{equation}
    \eta(v^2) = \frac{1}{nk-\frac{1}{2} k(k+1)} \int_{\St_{n,k}} \dist^2(\mathbb{E}[X], x) p_{\pr(X)}(x) \vol_{\St_{n,k}}(x),
\end{equation}
which is the true scalar variance of the random variable $\pr(X)$ corresponding to the sample scalar variance in \eqref{eq:scalarcariancesample}. Due to \cite{figueras2024parameter}, in the case of $\St_{3,1}=\mathbb{S}^2$, $\eta$ is known to be injective and is explicitly given by
\begin{equation}
\begin{aligned}    
\eta(\sigma^2):= \frac{1}{(2\pi)^{1/2}} \exp(-\frac{1}{2\sigma^2}) \int_0^\pi&  \phi^2\left(\frac{\cos(\phi)}{\sigma} + \left( \frac{\cos^2(\phi)}{\sigma^2} + 1\right) \right.\\
& \left.\frac{\Phi(\frac{\cos(\phi)}{\sigma})}{\varphi(\frac{\cos(\phi)}{\sigma})} \right) \sin(\phi)\dd\phi.
\end{aligned} 
\label{eq:S2eta}
\end{equation}
As of today, there is no such explicit formula for a general Stiefel manifold since $p_{\pr(X)}$ is difficult to express for $\St_{n,k}$ when $k\geq 2$. Because of this, we shall need the following conjecture in order to perform parameter inference on the projected observations.

\begin{conjecture}
\label{con:varbij}
    Let $X\in \mathbb{R}^{n\times k}$ be a normal random variable with $\mathbb{E}[X]\in \St_{n,k}$ and covariance $\Cov(X)=\sigma^2 \id_{\mathbb{R}^{n\times k}}$. Then the projected normal random variable has inherent average $\mathbb{E}[X]$. Moreover, there exists an injective function $\eta: [0,\infty)\rightarrow [0,\infty)$ such that the inherent scalar variance is uniquely determined by $\eta(\sigma^2)$ for each $\sigma>0$.
\end{conjecture}

The pseudo-code in Algorithm \ref{alg:approxeta} help us to numerically estimate the function $\eta$ using the Monte Carlo method, in the case that $\eta$ is difficult to explicitly express. That Algorithm \ref{alg:approxeta} converges follows from \cite[Theorem 2.5]{figueras2024parameter}.

\begin{algorithm}[H]
\caption{An outline of how to estimate the function $\eta$ on $\St_{n,k}$}
\label{alg:approxeta}
\begin{algorithmic}[1]
    \State Given $\sigma^2 >0$ and $\mu = I_{n,k}$.
    \State Draw $L$ samples $\{X_i\}$ from $N(\mu,\sigma^2 \id_{\mathbb{R}^{n\times k}})$
    \State Compute $Z_i=\pr(X_i)$ for each $i$.
    \State Compute $V_i=\log_{\mu}(Z_i)$ for each $i$.
    \State Let $B_j$ be an orthonormal basis for $T_{\mu}\St_{n,k}$ and estimate
    \[
    \eta(\sigma^2) \approx \frac{1}{L} \frac{1}{nk-\frac{1}{2} k(k+1)}\sum_{i=1}^L  \sum_{j=1}^{(nk-\frac{1}{2} k(k+1))} \langle V_i,B_j\rangle^2
    \]
\end{algorithmic}
\end{algorithm}

\begin{figure}[h]
    \centering
    \includegraphics[width=0.8\linewidth]{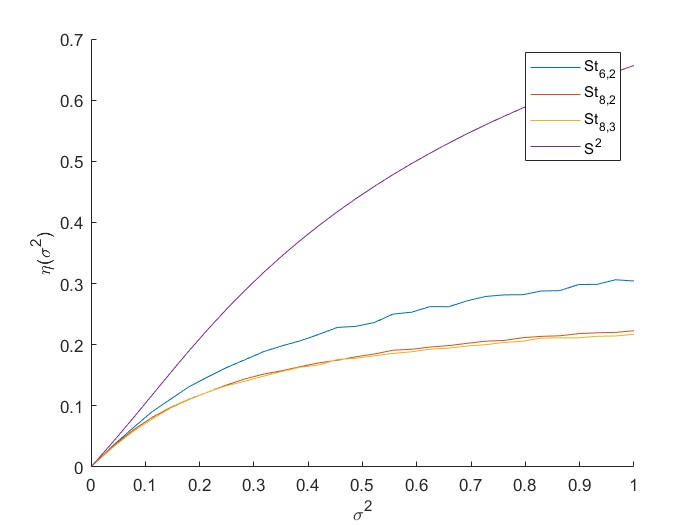}
    \caption{The intrinsic scalar variance of the projected normal as a function of the scalar variance of the original normal distribution.}
    \label{fig:etabij}
\end{figure}

Using Algorithm \ref{alg:approxeta}, we have numerically verified Conjecture \ref{con:varbij} for the Stiefel manifolds $\mathbb{S}^2$, $\St_{6,2}$, $\St_{8,2}$ and $\St_{8,3}$, see Figure \ref{fig:etabij}. For each $\sigma^2 \in (0,1]$ and each Stiefel manifold, 100 pseudo-random samples of $N(I_{n,k},\sigma^2 \id_{\mathbb{R}^{n\times k}})$ were drawn. Then, $\eta(\sigma^2)$ was estimated for each $\sigma^2$ and each Stiefel manifold. Notice that $\mathbb{S}^2$ is the only exception, in this case $\eta(\sigma^2)$ was computed by numerically approximating the integral in \eqref{eq:S2eta}. Notice also that each $\eta$ is roughly monotonically increasing and does not contradict Concjecture \ref{con:varbij}.

Now we have sufficient amount of tools to do filtering on $\St_{n,k}$.

\section{Extended Kalman Filtering on Stiefel Manifolds}
\label{sec:kalman}



Suppose we have the following filtering problem
\begin{equation}
    \begin{cases}
       \dd X_t  = A X_t \dd t + \nu \dd B_t, & X_0\eqdist N(\mu_0,\sigma^2_0 \id_{\mathbb{R}^{n\times k}})\\
       Z_m= \pr(\pr(X_{t_m})+\varepsilon_m)
    \end{cases},
    \label{eq:filteringSDE}
\end{equation}
where $t_m$ is an increasing sequences of discrete times, $\varepsilon_m\in T_{\pr(X_{t_m})} \St_{n,k}$ are i.i.d. such that $ \varepsilon_m \eqdist N(0,\xi^2 \id_{T_{\pr(X_{t_m})} \St_{n,k}})$, $\nu \in \mathbb{R}$ and $A:\mathbb{R}^n \rightarrow  \mathbb{R}^n$ is an anti-symmetric linear mapping ($A^T=-A$).

\begin{figure}[ht!]
    \centering
    \includegraphics[width=0.9\textwidth]{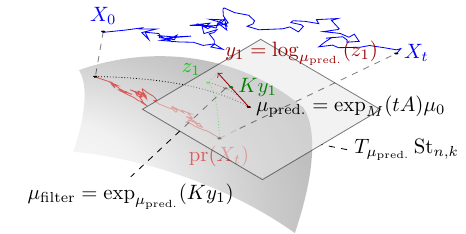}
    \caption{A geometrical picture of how the filtered average $\mu^K$ is computed after a timestep $t$ as a weighted average with weight Kalman gain $K$ between the predicted mean $\mu_{\text{pred.}}$ and the measurement $z_1$.}
    \label{fig:filteringstep}
\end{figure}

Next we shall adopt the classical Extended Kalman filter to the setting of Stiefel manifolds. Suppose we have prior $X_0\eqdist N(\mu_0,\sigma^2_0 \id_{\mathbb{R}^{n\times k}})$, and suppose after time $t_1$ we make measurement $z_1$. Now our predicted average is $\mu_{\text{pred.}}=\exp_{\mathrm{M}}(t_1A)\mu_0$ and the (inherent) predicted scalar variance is, by a linear approximation, $P_{\text{pred.}}=\eta(\sigma_0^2 + t_1\nu^2)$. Since the measurements have scalar variance $\xi^2$ we can compute the Kalman gain by
\[
K= \frac{\sigma_0^2 +t_1\nu^2}{\sigma_0^2 +t_1\nu^2+\xi^2}.
\]
We use the logarithm map to get the tangent vector $y_1 \in T_{\mu_{\text{pred.}}} \St_{n,k}$ which gives the velocity of the constant speed curve starting at $\mu_{\text{pred.}}$ which reaches $z_1$, i.e. $\exp_{\mu_{\text{pred.}}}(y_1)= z_1$. Now we rescale this vector by the Kalman gain $K$ to get a weighted average of the predicted point $\mu_{\text{pred.}}$ and the measurement $z_1$. We end up with the mean estimate $\exp_{\mu_{\text{pred.}}}(Ky_1)$ and the estimate $(1-K)P_{\text{pred.}}$ for the inherent (scalar) variance. Hence our posterior, after lifting this distribution back to the ambient space $\mathbb{R}^{n\times k}$, is
\[
N\left(\exp_{\mu_{\text{pred.}}}(Ky_1), \eta^{-1}((1-K)P_{\text{pred.}}) \id_{\mathbb{R}^{n\times k}}\right).
\]
In Figure \ref{fig:filteringstep} we give a picture of the geometry behind the filtering. Moreover, the entire extended Kalman filtering algorithm is given step-by-step in Algorithm \ref{alg:kalmansphere} as pseudo-code.

\begin{algorithm}[H]

\caption{One step of the extended Kalman filter on a manifold after time $t$}
\label{alg:kalmansphere}
\begin{algorithmic}[1]
    \State Given prior $x_0 \eqdist N(\mu_0,\sigma_0^2\id_{\mathbb{R}^{n \times k}})$ in $\mathbb{R}^{n \times k}$ with $\mu_0 \in \St_{n,k}$ and a measurement $z_1\in \St_{n,k} $.
    \State $F_{t} = \exp_{\mathrm{M}}(t A)$
    \State $\mu_{\text{pred.}} = F_t \cdot \mu_0$
    \State $P_{\text{pred.}} = \eta(\sigma_0^2 + t\nu^2)$
    \State $y= \log_{\mu_{\text{pred.}}}(z_1)$
    \State $S = \sigma_0^2 +t\nu^2 +\xi^2$
    \State $K= \left(\sigma_0^2 +t\nu^2 \right) S^{-1} $
    \State $\mu^K = \exp_{\mu_{\text{pred.}}}(K y)$
    \State $P^K = (1-K)P_{\text{pred.}}$
    \State The estimated Projected distribution is now $PrN(\mu^K,P^K \id_{T_{\mu^K} \St_{n,k}})$ with corresponding Normal distribution $N(\mu^K , \eta^{-1}(P^K)\id_{\mathbb{R}^{n \times k}}) )$
\end{algorithmic}
\end{algorithm}

\section[Simulations on the sphere]{Simulations on $\mathbb{S}^2$}
\label{sec:SimS2}

Throughout we consider system parameters 
\begin{equation}
    A= \begin{pmatrix}
        0 & 0.263& 0.036\\-0.263 & 0 & -0.653\\ -0.036& 0.653&0
    \end{pmatrix},
    \label{eq:simsystemA}
\end{equation}
$\nu^2 = 1$,  and $\xi^2 = 0.1$ with initial covariance $\Sigma_0=0.1 I_3$ and initial mean
\begin{equation}
    \mu_0=\begin{pmatrix}
        0\\ 0\\1
    \end{pmatrix}.
    \label{eq:siminitalmu}
\end{equation}

\begin{figure}[h]
    \centering
        \includegraphics[width=0.9\textwidth]{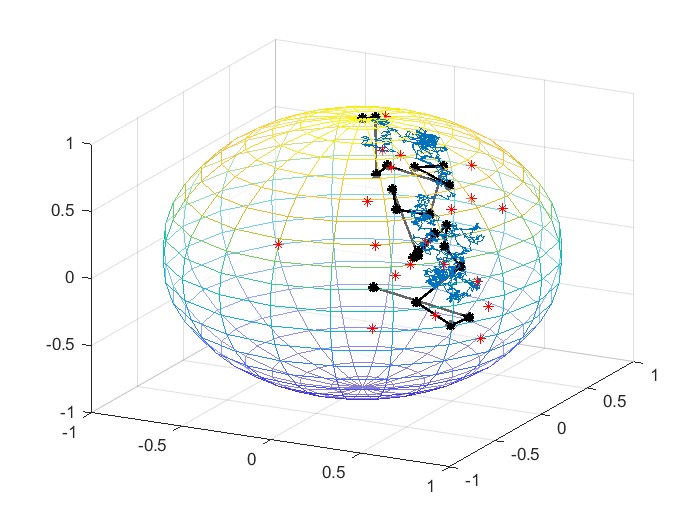}
        \caption{In blue the (projected) underlying stochastic process $X_t$, in red measurements $Z_{m}$ and in black the mean filtered estimate, with parameters $A$ given in \eqref{eq:simsystemA}, $\nu^2 = 1$, and $\xi^2 = 0.1$. The initial distribution is given by  $\Sigma_0=0.1 I_3$ and $\mu_0$ is given by \eqref{eq:siminitalmu}.}
        \label{fig:SimulatedKalmanfiltersphere}
\end{figure}

Now consider stochastic process $X_t$ solving \eqref{eq:systemsde}. The system process $X_t$ is simulated as follows: Consider constant interval discrete times $0=t_0,\dots t_N$ where $\Delta t= t_j-t_{j-1}$ for $1\leq j \leq N$. We choose $N=2000$. The initial point is drawn from $X_0= N(\mu_0,\Sigma_0 )$. Now each point $X_{t_j}$ is simulated by drawing $V_j$ from $N(0, \Delta t \nu^2I_{3})$ and computing 
\[
X_{t_j} = \exp_{\mathrm{M}}(\Delta t A) X_{t_{j-1}} + V_j.
\]
The projected system process $\pr(X_t)=\frac{X_t}{\norm{X_t}}$ can be seen in Figure \ref{fig:SimulatedKalmanfiltersphere} in blue.

At even time intervals $\tau_1,\dots, \tau_L=1$ with $L=20$ we simulate measurements of $X_{\tau_j}$ as follows: For each $\tau_j$ we pseudo-randomly draw $W_j$ from $ N(0,\xi^2 I_3)$ and then compute $Z_{\tau_j}=\pr(\pr(X_{\tau_j})+W_j)$. These measurements can be seen in Figure \ref{fig:SimulatedKalmanfiltersphere} in red. The predictions filtered that arise from the application of Algorithm \ref{alg:kalmansphere} to the measurements $Z_{\tau_j}$  are visualized in Figure \ref{fig:SimulatedKalmanfiltersphere} in black.

\begin{figure}[h]
        \centering
    \begin{subfigure}{.32\textwidth}
        \includegraphics[width=\textwidth]{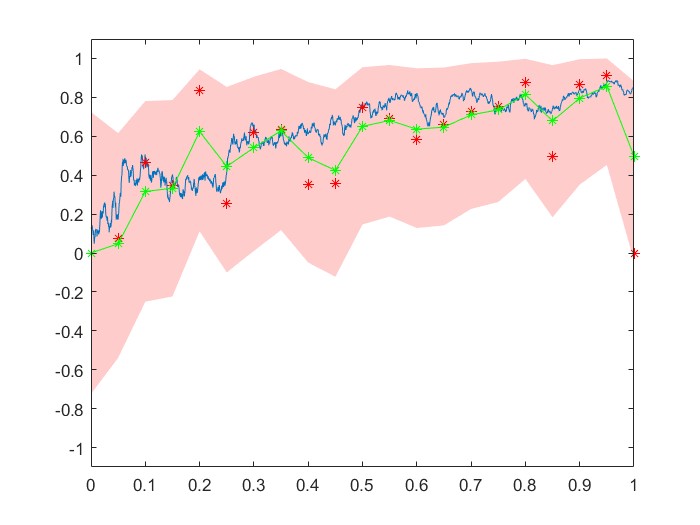}
        \caption{The $x$-coordinate.}
        \label{fig:xaxis}
    \end{subfigure}
    \begin{subfigure}{.32\textwidth}
        \includegraphics[width=\textwidth]{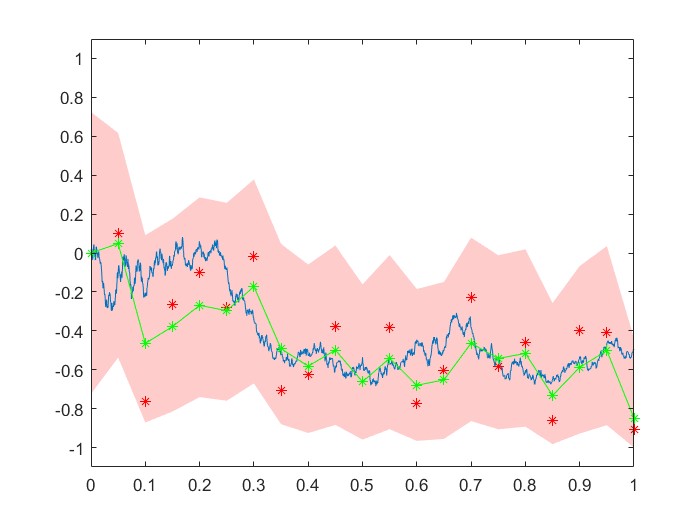}
        \caption{The $y$-coordinate.}
        \label{fig:yaxis}
    \end{subfigure}
    \begin{subfigure}{.32\textwidth}
        \includegraphics[width=\textwidth]{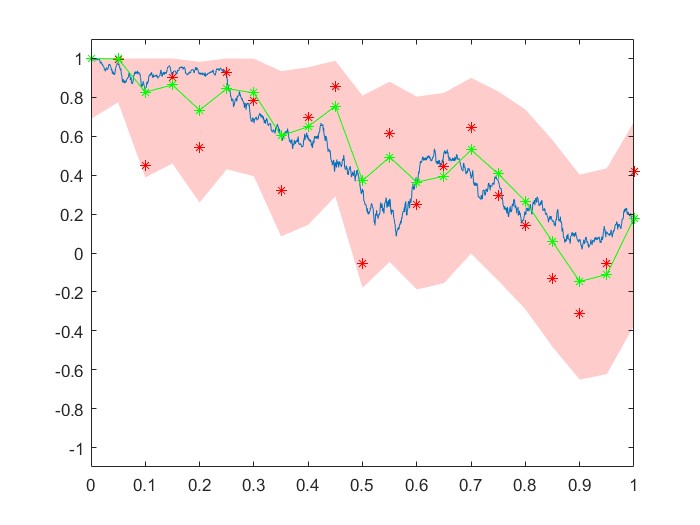}
        \caption{The $z$-coordinate.}
        \label{fig:zaxis}
    \end{subfigure}
    \caption{Simulated and filtered data over a time span of $1$ time unit. In blue the (projected) underlying stochastic process $X_t$, in red, measurements $Z_{m}$, and in green, the mean filtered estimate. The pink filled regions represent a $95\%$-confidence intervals of the filtered estimate. To run this simulation the parameters $A$ given in \eqref{eq:simsystemA}, $\nu^2 = 1$, and $\xi^2 = 0.1$  were used. The initial distribution is given by  $\Sigma_0=0.1 I_3$ and $\mu_0$ is given by \eqref{eq:siminitalmu}.}
\end{figure}

 To illuminate this simulation into something more easily readable, we have also provided Figure \ref{fig:xaxis} which shows the $x-$coordinate of the same data as in Figure \ref{fig:SimulatedKalmanfiltersphere} over time but including a $95\%$-confidence interval around the filtered mean. Similarly, the $y$-coordinate is shown in Figure \ref{fig:yaxis} and the $z$-coordinate in Figure \ref{fig:zaxis}.

As can be seen in Figures \ref{fig:SimulatedKalmanfiltersphere}, \ref{fig:xaxis}, \ref{fig:yaxis} and \ref{fig:zaxis}, the filtered mean is on average closer to the underlying process compared to the distance from the measurements to the underlying process. This tendency will be made clearer in the next set of simulations. We redo the simulation above for $\nu^2=0.1,0.2,0.5,1.0$. Then, for each such $\nu^2$ we simulate the filter with 100-fold repetitions for different choices of $\xi^2$. This is done for 
\[
\xi^2= \nu^2,\frac{\nu^2}{2.8},\frac{\nu^2}{4.6},\frac{\nu^2}{6.4},\frac{\nu^2}{8.2}, \frac{\nu^2}{10}.
\]
In \Cref{fig:SNRnu01S2,fig:SNRnu05S2,fig:SNRnu02S2,fig:SNRnu10S2}, we plot, for different signal to noise ratios in units decibel, i.e $10 \log_{10}\left(\frac{\eta(\nu^2)}{\xi^2} \right)$ (dB), the average distance from true point $X_t$ to the measurements, and to the filtered means, respectively. As can be seen for all Figures in Figure \ref{fig:SNRS2} there is a significant improvement on the average error of the filtered mean compared to the error of the raw measurements.

\clearpage
\begin{figure}[H]
    \centering
    \begin{subfigure}{0.8\textwidth}
        \includegraphics[width=\textwidth]{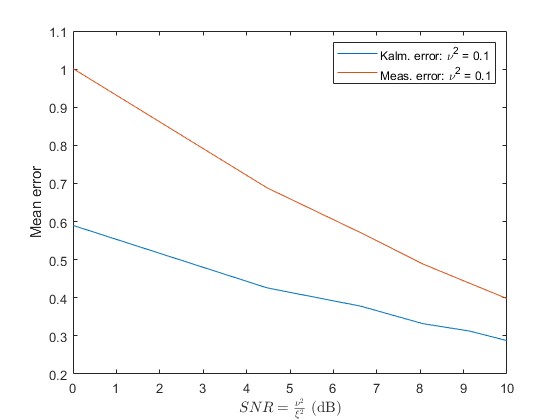}
        \caption{SNR plot for $\nu^2=0.1$.}
        \label{fig:SNRnu01S2}    
    \end{subfigure}
\end{figure}
\begin{figure}[H]
    \ContinuedFloat
    \centering
    \begin{subfigure}{0.8\textwidth}
        \includegraphics[width=\textwidth]{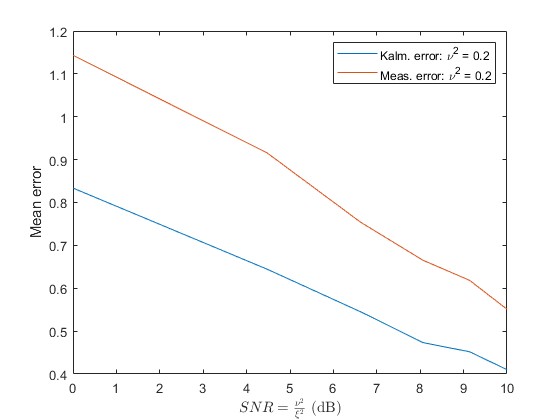}
        \caption{SNR plot for $\nu^2=0.2$.}
        \label{fig:SNRnu02S2}    
    \end{subfigure}
\end{figure}
\begin{figure}[H]
    \ContinuedFloat
    \centering
    \begin{subfigure}{0.8\textwidth}
        \includegraphics[width=\textwidth]{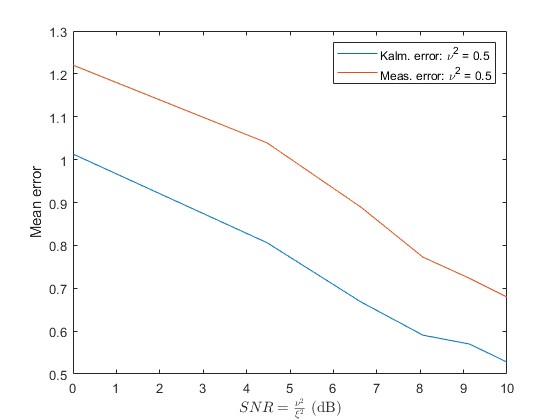}
        \caption{SNR plot for $\nu^2=0.5$.}
        \label{fig:SNRnu05S2}    
    \end{subfigure}
\end{figure}
\begin{figure}[H]
    \ContinuedFloat
    \centering
    \begin{subfigure}{0.8\textwidth}
        \includegraphics[width=\textwidth]{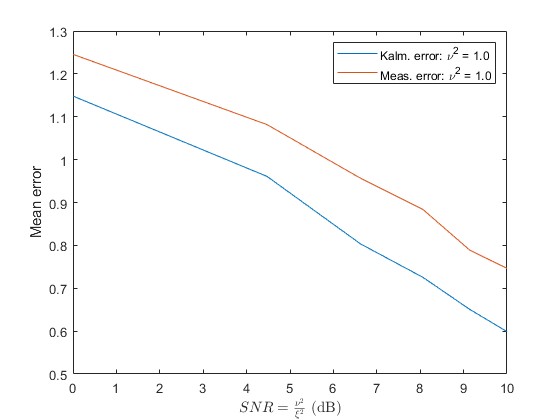}
        \caption{SNR plot for $\nu^2=1.0$.}
        \label{fig:SNRnu10S2}    
    \end{subfigure}
    \caption{For different SNR defined by $\nu^2/\xi^2$ given different values of $\nu^2$ the following is plotted: The average error of the filtered mean is the average distance from $\mu^K_m$ to $X_{t_m}$ in blue and the average error of the measurements is the average distance from $Z_{m}$ to $X_{t_m}$ in red. The errors are produced by averaging over 100 realisations of filtering over 20 measurements for each set of parameters $\nu,\xi$.}
    \label{fig:SNRS2}
\end{figure}

\begin{table}[H]
\caption{Tabulated data corresponding to Figure \ref{fig:SNRS2}.}
\label{tab:SNRS2}
\begin{tabular}{|l|l|l|l|l|l|l|l|}
\hline
\multicolumn{1}{|l|}{$\nu^2$} & SNR (dB)     & 0    & 4.47 & 6.63 & 8.06 & 9.14 & 10  \\ \hline
\multirow{2}{*}{0.1}          & Error meas.  & 1.00 & 0.69 & 0.57 & 0.49 & 0.44 & 0.40 \\ \cline{2-8} 
                              & Error filter & 0.59 & 0.43 & 0.38 & 0.33 & 0.31 & 0.29 \\ \hline
\multirow{2}{*}{0.2}          & Error meas.  & 1.14 & 0.92 & 0.75 & 0.67 & 0.62 & 0.55 \\ \cline{2-8} 
                              & Error filter & 0.83 & 0.64 & 0.55 & 0.47 & 0.45 & 0.41 \\ \hline
\multirow{2}{*}{0.5}          & Error meas.  & 1.22 & 1.04 & 0.89 & 0.77 & 0.72 & 0.68 \\ \cline{2-8} 
                              & Error filter & 1.01 & 0.81 & 0.67 & 0.59 & 0.57 & 0.53 \\ \hline
\multirow{2}{*}{1.0}          & Error meas.  & 1.25 & 1.08 & 0.96 & 0.88 & 0.79 & 0.75 \\ \cline{2-8} 
                              & Error filter & 1.15 & 0.96 & 0.80 & 0.73 & 0.65 & 0.60 \\ \hline
\end{tabular}
\end{table}

\section[Simulations on St]{Simulations on $\St_{4,2}$}
\label{sec:simSt}

Here we consider the filtering problem in \eqref{eq:filteringSDE} on $\St_{4,2}$. We choose system parameters
\begin{equation}
    A= \begin{pmatrix}
        0 & 0.173& 0.267 & -0.288\\-0.173 & 0 & -0.279 & 0.122\\ -0.267& 0.279&0 & 0.316\\ 0.288 & -0.122 & -0.316& 0
    \end{pmatrix}
    \label{eq:simsystASt}
\end{equation}
$\nu^2 = 1$ and $\xi^2 =0.1$, with initial mean
\begin{equation}
\mu_0=\begin{pmatrix}
        1& 0\\0&1\\0 &0\\0&0
    \end{pmatrix},
    \label{eq:siminitalmuSt}
\end{equation}
and initial covariance $\Sigma_0=0.1 \id_{\mathbb{R}^{4\times 2}}$.

The system process $X_t\in \mathbb{R}^{4,2}$ is done equivalently to how it is done for $\mathbb{S}^2$ in the previous section. Each coordinate of the projected system process $\pr(X_t)= X_t(X_t^T X_t)^{-1/2}$ can be seen in Figure \ref{fig:simulationstiefel} in blue.

Similarly to before, given even time intervals $\tau_1,\dots, \tau_L=1$ we simulate measurements of $X_{\tau_j}$ as follows: For each $\tau_j$ we pseudo-randomly draw $W_j$ from $ N(0.\xi^2 \id_{\mathbb{R}^{4\times 2}})$ then we compute $Z_{\tau_j}=\pr(\pr(X_{\tau_j})+W_j)$, here $L=20$. These measurements can be seen in red in Figure \ref{fig:simulationstiefel}. By directly applying Algorithm \ref{alg:kalmansphere} with the measurements $Z_{\tau_j}$ the filtered predictions is visualized in green in Figure \ref{fig:simulationstiefel}. Note that here we roughly give $95$-\% confidence bounds as intervals of the form $[\mu^K-1.96 \sqrt{P^K},\mu^K-1.96 \sqrt{P^K}]$.

\begin{figure}[H]
    \centering
    \begin{subfigure}{.32\textwidth}
        \includegraphics[width=\textwidth]{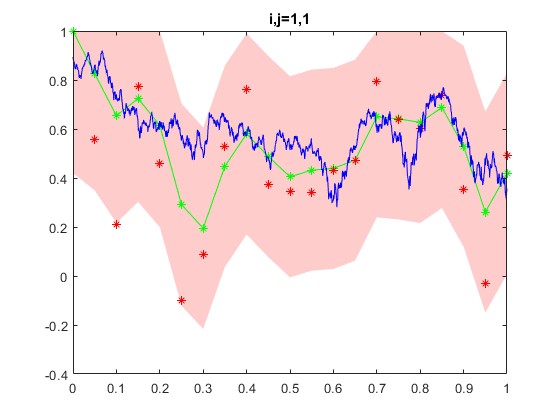}
        \caption{Coordinate $X_{11}$.}
        \label{fig:stiefel11}
    \end{subfigure}
    \begin{subfigure}{.32\textwidth}
        \includegraphics[width=\textwidth]{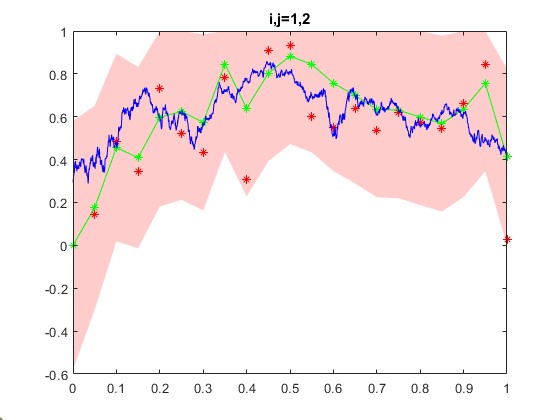}
        \caption{Coordinate $X_{12}$.}
        \label{fig:stiefel12}
    \end{subfigure}
    \begin{subfigure}{.32\textwidth}
        \includegraphics[width=\textwidth]{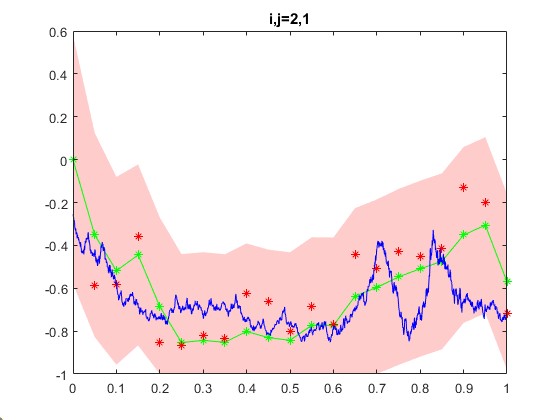}
        \caption{Coordinate $X_{21}$.}
        \label{fig:stiefel21}
    \end{subfigure}
\end{figure}
\begin{figure}[H]
    \ContinuedFloat
    \centering
    \begin{subfigure}{.32\textwidth}
        \includegraphics[width=\textwidth]{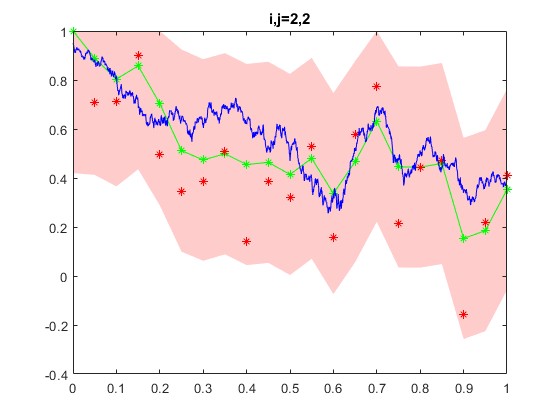}
        \caption{Coordinate $X_{22}$.}
        \label{fig:stiefel22}
    \end{subfigure}
    \begin{subfigure}{.32\textwidth}
        \includegraphics[width=\textwidth]{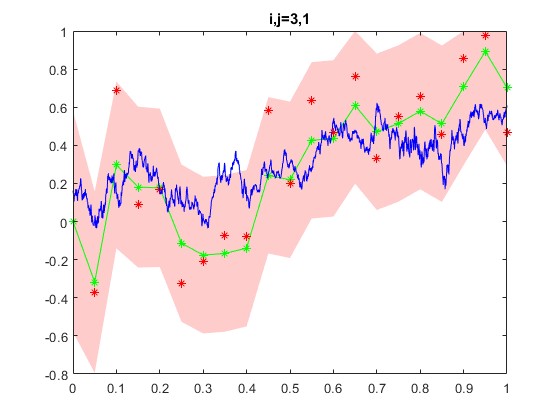}
        \caption{Coordinate $X_{31}$.}
        \label{fig:stiefel31}
    \end{subfigure}
    \begin{subfigure}{.32\textwidth}
        \includegraphics[width=\textwidth]{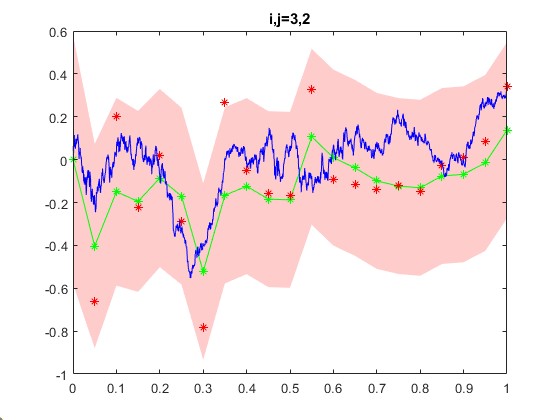}
        \caption{Coordinate $X_{32}$.}
        \label{fig:stiefel32}
    \end{subfigure}
\end{figure}
\begin{figure}[H]
    \ContinuedFloat
    \centering
    \begin{subfigure}{.32\textwidth}
        \includegraphics[width=\textwidth]{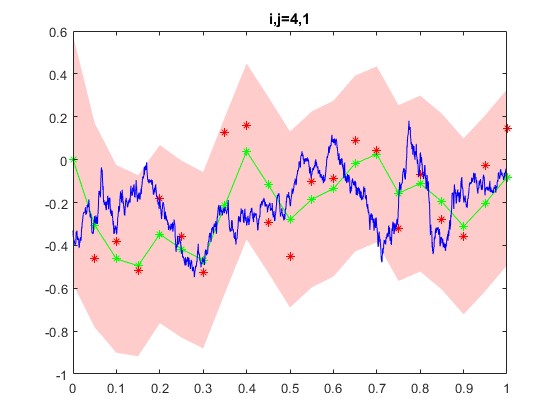}
        \caption{Coordinate $X_{41}$.}
        \label{fig:stiefel41}
    \end{subfigure}
    \begin{subfigure}{.32\textwidth}
        \includegraphics[width=\textwidth]{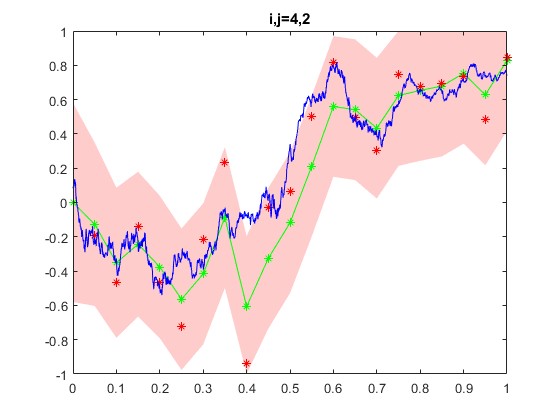}
        \caption{Coordinate $X_{42}$.}
        \label{fig:stiefel42}
    \end{subfigure}
\end{figure}
\begin{figure}[H]
    \ContinuedFloat
    \centering
    \caption{Simulated and filtered data over a time span of $1$ unit of time. Throughout, the (projected) underlying stochastic process $X_t$ is depicted in blue, the measurements are depicted in red, the filtered averages are depicted in green together with pink filled regions representing an approximated $95\%$-confidence interval for each filtered mean estimate. The parameter $A$ is given by \eqref{eq:simsystASt}, $\nu^2 = 1$, and $\xi^2 =0.1$. The initial mean $\mu_0$ is given by \eqref{eq:siminitalmuSt} and the initial covariance is $\Sigma_0^2=0.1\id_{\mathbb{R}^{4\times 2}}$.}
    \label{fig:simulationstiefel}
\end{figure}

As can be seen in Figure \ref{fig:simulationstiefel}, on average, the filtered mean is closer to the underlying process compared to the measurements. Again, we shall verify this in the following simulations. We once more define SNR as $10 \log_{10}\left(\frac{\nu^2}{\xi^2} \right)$ (dB). In \Cref{fig:SNRnu01,fig:SNRnu02,fig:SNRnu05,fig:SNRnu10} one can see the average error of the filtered mean and the average error of the measurements. This is done with the same choices of $\nu$ and $\xi$ as was done for $\mathbb{S}^2$ in the previous section. As can be seen in \Cref{fig:SNRnu01,fig:SNRnu02,fig:SNRnu05,fig:SNRnu10}, the error of the filtered mean is significantly smaller compared to the error of the measurements.

\begin{figure}[H]
    \centering
    \begin{subfigure}{0.8\textwidth}
        \includegraphics[width=\textwidth]{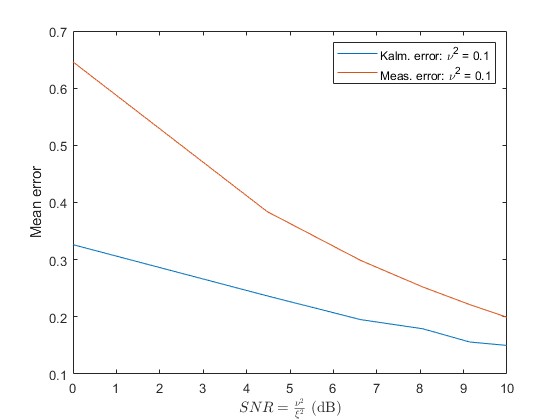}
        \caption{SNR plot for $\nu^2=0.1$.}
        \label{fig:SNRnu01}    
    \end{subfigure}
\end{figure}
\begin{figure}[H]
    \ContinuedFloat
    \centering
    \begin{subfigure}{0.8\textwidth}
        \includegraphics[width=\textwidth]{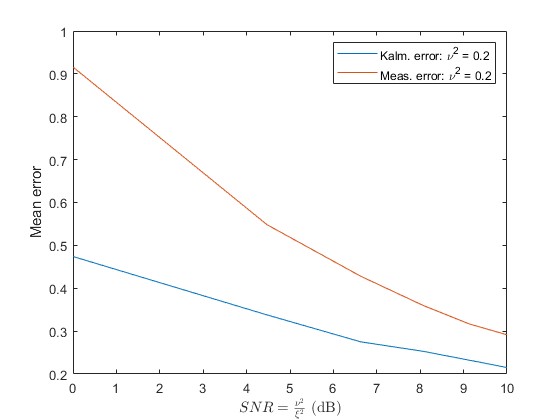}
        \caption{SNR plot for $\nu^2=0.2$.}
        \label{fig:SNRnu02}    
    \end{subfigure}
\end{figure}
\begin{figure}[H]
    \ContinuedFloat
    \centering  
    \begin{subfigure}{0.8\textwidth}
        \includegraphics[width=\textwidth]{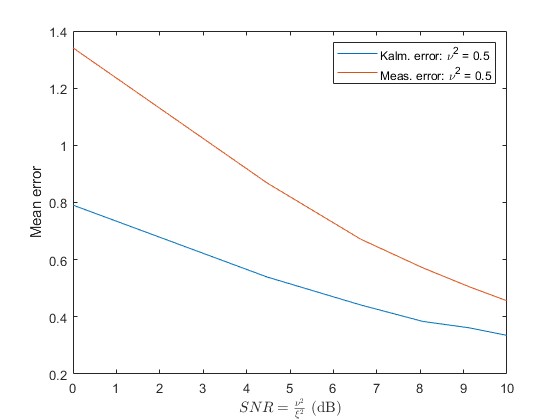}
        \caption{SNR plot for $\nu^2=0.5$.}
        \label{fig:SNRnu05}    
    \end{subfigure}
\end{figure}
\begin{figure}[H]
    \ContinuedFloat
    \centering
    \begin{subfigure}{0.8\textwidth}
        \includegraphics[width=\textwidth]{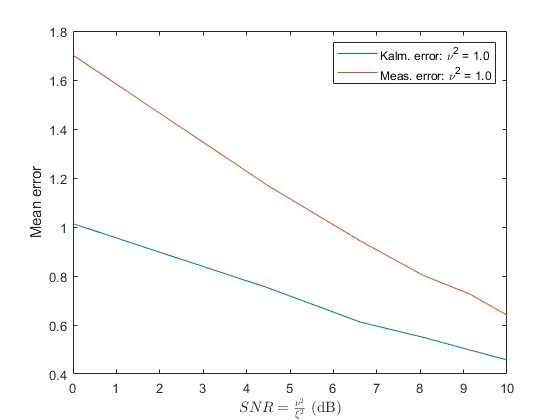}
        \caption{SNR plot for $\nu^2=1.0$.}
        \label{fig:SNRnu10}    
    \end{subfigure}
    \caption{For different SNR defined by $\nu^2/\xi^2$ given different values of $\nu^2$ the following is plotted: The average error of the filtered mean is the average distance from $\mu^K_m$ to $X_{t_m}$ in blue. The average error of the measurements is the average distance from $Z_{m}$ to $X_{t_m}$ in red. The errors are produced by averaging over 100 realisations of filtering over 20 measurements for each set of parameters $\nu,\xi$.}
    \label{fig:SNRstiefel}
\end{figure}


\begin{table}[H]
\caption{Tabulated data corresponding to Figure \ref{fig:SNRstiefel}.}
\label{tab:SNRSt}
\begin{tabular}{|l|l|l|l|l|l|l|l|}
\hline
$\nu^2$              & SNR (dB)     & 0    & 4.47 & 6.63 & 8.06 & 9.14 & 10   \\ \hline
\multirow{2}{*}{0.1} & Error meas.  & 0.65 & 0.39 & 0.30 & 0.25 & 0.22 & 0.21 \\ \cline{2-8} 
                     & Error filter & 0.33 & 0.23 & 0.20 & 0.18 & 0.16 & 0.15 \\ \hline
\multirow{2}{*}{0.2} & Error meas.  & 0.92 & 0.54 & 0.42 & 0.36 & 0.32 & 0.29 \\ \cline{2-8} 
                     & Error filter & 0.46 & 0.33 & 0.28 & 0.24 & 0.22 & 0.21 \\ \hline
\multirow{2}{*}{0.5} & Error meas.  & 1.35 & 0.87 & 0.68 & 0.57 & 0.51 & 0.45 \\ \cline{2-8} 
                     & Error filter & 0.74 & 0.54 & 0.44 & 0.38 & 0.36 & 0.33 \\ \hline
\multirow{2}{*}{1.0} & Error meas.  & 1.65 & 1.20 & 0.96 & 0.82 & 0.72 & 0.65 \\ \cline{2-8} 
                     & Error filter & 1.09 & 0.74 & 0.62 & 0.56 & 0.50 & 0.47 \\ \hline
\end{tabular}%
\end{table}

\section{Discussion}
\label{sec:discussion}

As can be seen in  Figures \ref{fig:SNRS2} and \ref{fig:SNRstiefel} depending on the situation, one can expect at optimal conditions about a 60\% filtering gain. Note, however, that if $\nu$ is too large, then the projected distribution will be close to the uniform distribution. In this case the extended Kalman filtering algorithm is not feasible as the predictive power of the dynamics becomes unreliable.
This is also true for the error of the measurements, if the error of the measurements are large. Indeed, then there is little information gained from the measurements, naturally. 

Some inaccuracies might originate from the logarithm map used in step 5 of Algorithm \ref{alg:kalmansphere}. The logarithm map is only defined for points close to the base point, therefore one can not guarantee that the algorithm is reliable if the measurements appear outside what is called the injectivity radius (the radius which guarantees that the logarithm map is well defined). It is still an open question what the injectivity radius is for the Stiefel manifold. However, very recently there has been some advances in the following direction: there is a lower bound $\sqrt{\frac{4}{5}}\pi$ for the injectivity radius of the Stiefel manifold (under the canonical metric), see \cite{stoye2024injectivity}. Therefore, as long as points are never farther away from each other than $\sqrt{\frac{4}{5}}\pi$, the injectivity of the logarithm map is not an issue.

One obvious limitation with Algorithm \ref{alg:kalmansphere} is that it requires quite restrictive assumptions on the noise, i.e. we require that the noise is isotropic. Indeed, already on $\mathbb{S}^2$, if we allow for general covariance matrices $\Sigma$ then the statement of Conjecture \ref{con:varbij} is rarely true. Thus, if the filtering problem is set up so that future points may not allow for one-to-one correspondence between the covariance of a normal random variable and it's projected counterpart, our proposed filter is useless. In the case of the sphere, as long as the covariance matrix $\Sigma$ of the ambient normal random variable has the mean as eigenvector, then Algorithm \ref{alg:kalmansphere} may still be tractable.

One more subtle issue with Algorithm \ref{alg:kalmansphere} is that the update steps are only accurate as long as the actual process $X_t$ does not significantly drift away from it's starting point in the radial direction. By choosing $A$ to be anti-symmetric, $\exp_{\mathrm{M}}(tA)\in SO(n)$, then it is always true that $\exp_{\mathrm{M}}(tA) \mu_0\in \St_{n,k}$ if $\mu_0\in \St_{n,k}$. Therefore on average, the process $X_t$ does not drift away along a radial direction. However, that does not mean that it's realisation doesn't drift away. In order to bootstrap the filtering process, one way periodically measure the process noise term $\eta(\nu^2)$, of $\pr(X_t)$, since $\eta(\nu^2)$ also depends on the symmetric part $S$ of the polar decomposition $X_t = \pr(X_t) S$.

\printbibliography

\end{document}